\documentclass{article}
\usepackage{a4}
\usepackage{amsfonts}
\usepackage{t1enc}
\usepackage[dvips]{epsfig}
\usepackage[latin1]{inputenc}

\begin{document}

\newcommand{\ord}{\ensuremath{\mbox{ \rm ord}}}
\newcommand{\cfrac}{\ensuremath{\mbox{ \rm cfrac}}}
\newcommand{\lcm}{\ensuremath{\mbox{ \rm lcm}}}
\newtheorem{definition}{Definition}
% erste {}: Name der Struktur latexintern
% zweite {}: Name der Struktur im ausgedruckten Teil
% []: Orientierung der Numerierung (hier an \section)

\newtheorem{theorem}[definition]{Theorem}
% [definition] git an, dass die Theoreme und die Definitionen einheitlichen Zaehler haben (1.1. Def, 1.2. Theor...)
% etc
\newtheorem{lemma}[definition]{Lemma}
\newtheorem{korollar}[definition]{Corollary}
\newtheorem{bemerkung}[definition]{Bemerkung}
\newtheorem{beispiel}[definition]{Beispiel}
\newtheorem{algo}[definition]{Algorithmus}
\newtheorem{postulat}{Postulat}

\title{Improving the Success Probability for Shor`s Factoring Algorithm}
\author{Gregor Leander\\Institute for IT-Security and
Cryptology\\Ruhr Universität Bochum}
\maketitle

\abstract{}
Given $n=pq \in \mathbb{N}$ with $p$ and $q$ prime and $y \in \mathbb{Z}_n^*$ . Shor's algorithm
computes the order of $y$.
\[ y^r = 1 \pmod{n} ,\]
and if $r=2k$, we get
\[ (y^k-1)(y^k+1)= 0 \pmod{n} .\]
Assuming that $y^k \ne -1 \pmod{n}$, we can easily compute a non
trivial factor of $n$:
\[ \gcd(y^k-1,n).\]
In \cite{ShorAlgo} it is shown that a randomly chosen $y$ is usable for
factoring with probability at least $\frac{1}{2}$.
In this paper we will show an efficient possibility to improve the lower bound of
this probability by selecting only special $y\in \mathbb{Z}_n^*$. The
lower bound of the probability using only this $y$ as an input for
Shor`s algorithm is
$\frac{3}{4}$, so we have reduced the fault probability in the worst
case from $\frac{1}{2}$ to $\frac{1}{4}$. 
\section*{Preprocessing for Shor`s Algorithm}

The following lemma is the starting point of our discussion: 
\begin{lemma}\label{WktFuerGuteOrdnung}\label{BedinungenFuerFaktor}
Let $n=pq$ with $p,q$ prime. Then at least half of the $y \in
\mathbb{Z}_n^*$ satisfy the following conditions:
\begin{eqnarray}
	\mbox{ The order } r \mbox{ of } y \mbox{ is even, i.e. } \exists k \mbox{ with } r=2k \label{one}\\
	y^k\ne -1 \pmod{n} \label{two}
\end{eqnarray}\hfill
\end{lemma}
If $p-1=2^ns$ and $q-1=2^ms$, with $m,s$ odd, the probability is
exactly given by
\[2^{-(m+n)}  \left(1+ \sum_{j=0}^{min\{m,n\}-1}4^j \right)
\le\frac{1}{2}.\]

The easy but helpful observation in order to improve this lower bound
is the following lemma. 

\begin{lemma}
Let $p$ be a prime and $a$ a non-square in $\mathbb{Z}_p^*$.
Then the order of $a$ is even.
\end{lemma}
{\bf Proof:} Let $g$ be a generator of $\mathbb{Z}_p^*$ and
$a=g^s$. As $a$ is a non-square it follows that $s$ is odd. The order
of a satisfies $\ord_p(a)s=0 \bmod{p-1}$ i.e. $\ord_p(a)s=k(p-1)$, and
that means that $\ord_p(a)$ has to be even.
\hfill 

An element $y$ in  $\mathbb{Z}_n^*$ has even order, if $y$ has even
order in $\mathbb{Z}_p^*$ or $\mathbb{Z}_q^*$. This yields to the
following corrolary:
\begin{korollar}\label{EvenOrder} Let $y$ be any element in $\mathbb{Z}_n^*$. Then
\[ \left( \frac{y}{n} \right)=-1 \Rightarrow \exists k \mbox{ such that
}\ord_n(y)=2k\]
\hfill 
\end{korollar}
As the Jacobi-Symbol is efficiently computable, we now have a
sufficient criterion for an element to have even order.

Putting this together with the condition that $y^k \ne -1$ we get our
main theorem:

\begin{theorem}\label{NEW}
The probability that a random $y \in  \mathbb{Z}_n^*$ with  $\left(
\frac{y}{n} \right)=-1$ satisfies 
\[ \ord_n(y)=2k \mbox{ and } y^k \ne -1 \pmod{n} \] 
is at least $\frac{3}{4}$.
\end{theorem}

To proof the theorem we need:
\begin{lemma}\label{OrdnungNachPotenz} Let $p$ be prime with $p-1=2^mx$,
$x$ odd. Further let $g$ be a generator of $\mathbb{Z}_p^*$ and $b \in \mathbb{Z}_p^*$.
\begin{enumerate}
	\item For $k \in \{1,...,m\}$: 
	\[ \ord_n(b)=2^kw \mbox{,w odd } \Leftrightarrow
	b=g^{2^{(m-k)}v} \mbox {,v odd } \]
	In particular there are $2^{k-1}x$ elements of this form. 
	\item The order of $b$ is odd, if and only if $b= g^{2^mw}$
	with $1 \le w \le x$ in $\mathbb{Z}_p^*$. There are exactly
	$x$ elements with odd order.
\end{enumerate}
\end{lemma}
{\bf Proof:} 
\begin{enumerate}
	\item Let $b=g^s$ and $t$ be the order of $b$. This is
	equivalent with $st=0 \bmod{(p-1)}$, $t$ minimal. That means, 
\[ t=\frac{p-1}{\gcd(p-1,s)} .\]
If $t$ is of the form $2^kw$ ($w$ odd),  $2^{m-k}$ divides $s$
	but $2^{m-k+1}$ does not.This proves the statement.
	\item The order $t$ is odd, iff $2^m$ divides $s$, and this
	means that $s$ is $s=2^mw$ with $1 \le w \le x$.
\end{enumerate}
 \hfill

{\bf Proof of the theorem:} We are going to count the elements  $y\in  \mathbb{Z}_n^*$
with $\left( \frac{y}{n} \right)=-1$ not
satisfying the condition (\ref{two}). Due to corollary \ref{EvenOrder} we
know that the order of $y$ in $\mathbb{Z}_n^*$ is
even, i.e. $\ord_n(y)=2k$. We denote $s=\ord_p(y)=2^iv$ and
$t=\ord_q(y)=2^jw$ with $v,w$ odd. In particular
$2k=2^{\max\{i,j\}}\lcm(v,w)$. The $y$ we are counting fulfill $y^k=-1
\bmod{n}$ and this is equivalent to $y^k=-1 \bmod{p}$ and $y^k=-1
\bmod{q}$. But this means that neither $s$ nor $t$ divides $k$ (because otherwise for example $y^k=y^{cs}=1 \bmod{p}$) and it follows that $i=j$.
$\left( \frac{y}{n} \right)=-1$ means, that $\left( \frac{y}{p}
 \right)=-1$ and $\left( \frac{y}{q} \right)=1$ or  $\left(
 \frac{y}{p} \right)=1$ and  $\left( \frac{y}{q} \right)=-1$.
W.l.o.g we assume the first case is true:

Let $p-1=2^{m_1}x_1$ and $g_p$ be a generator of $\mathbb{Z}_p^*$,
then $\left( \frac{y}{p} \right)=-1$ if and only if $y=g_p^{t_1}$ for
odd $t_1$. So we have to count all the odd $t_1$, such that the order of
$y=g_p^{t_1}$ is of the form $2^iv$, $v$ odd. With lemma
\ref{OrdnungNachPotenz} we conclude that only for $i=m_1$ such values $t_1$
can exist, and in this case all odd values between $1$ and $p-1$ lead
to such an element.

Now we have to discuss the elements with respect to $q$. Let
$q-1=2^{m_2}x_2$ and $g_q$ be a generator of $\mathbb{Z}_q^*$. We have
to count all the even values $t_2$ where the order of $g^{t_2}$ if of
the form $2^{m_1}w$. When $m_1 >m_2$, there are no such values
because the order of $g^{t_2}$ has to divide $q-1=2^{m_2}x_2$.If
$m_1=m_2$ there are no even solutions for $t_2$.
So the only case remaining is $m_1<m_2$. Due to lemma
\ref{OrdnungNachPotenz} the solutions are exactly the $t_2$ of the
form $2^{m_2-m_1}u$ for $u$ odd. Here $u$ can be any odd value between
$1$ and $2^{m_1}x_2-1$, so this gives exactly $2^{m_1-1}x_2$
solutions.

For the second case $\left( \frac{y}{p} \right)=1$ and  $\left(
\frac{y}{q} \right)=-1$ we get the same result, so in the case $m_1
\ne m_2$ we can assume $m_1 <m_2$ w.l.o.g.. 

Summing up all these values not satisfying the conditions  (\ref{one}) and (\ref{two}) when $m_1
\ne m_2$ we get:
\begin{eqnarray*}
	A(n)&=&\frac{p-1}{2}2^{(m_1-1)}x_2=\frac{p-1}{2} \frac{2^{m_2}x_2}{2^{(m_2-m_1+1)}} \\
	&=& \frac{1}{4}(p-1)(q-1) \frac{1}{2^{(m_2-m_1)}} \\
	&=&  \frac{1}{4} \varphi(n) \frac{1}{2^{(m_2-m_1)}}
\end{eqnarray*}
The number of elements with $\left( \frac{y}{n} \right)=-1$ is $
\frac{1}{2} \varphi(n)$ and so the probability we where looking for
is:
\[ P(n) =1-\frac{A(n)}{\varphi(n)} =1- \frac{1}{2^{(m_2-m_1+1)}} \ge \frac{3}{4} .\]
The case $m_1=m_2$ is even better, because here the probability is $1$
that means that $y$ with $\left( \frac{y}{n} \right)=-1$ always satisfies
both conditions (\ref{one}) and (\ref{two}).
\hfill

\end{document}